\documentstyle[12pt,amsfonts]{article}
\def\one{1\hskip-.37em 1}

\def\half{\textstyle{\frac{1}{2}}}

\def\H{{\cal H}}
\def\D{{\cal D}}

\def\ra{\rightarrow}
\def\tint{{\textstyle\int}}
\def\d{\partial}
\def\o{\overline}
\def\C{{\cal C}_\beta}
\def\a{\alpha}
\def\ep{\epsilon}
\def\b{\begin{eqnarray*}}     
\def\e{\end{eqnarray*}}       
\def\bn{\begin{eqnarray}}     
\def\en{\end{eqnarray}}       
\def\<{\langle}
\def\>{\rangle}

\def\{{\lbrace}
\def\}{\rbrace}
\begin{document}
\title{Coherent State Path Integrals {\it without} Resolutions of Unity}   
\author{John R. Klauder\\
Departments of Physics and Mathematics\\
University of Florida\\
Gainesville, FL  32611 \\
Email: klauder@phys.ufl.edu}
\date{}
\maketitle
\begin{abstract}
>From the very beginning, coherent state path integrals have always relied
on a coherent state resolution of unity for their construction. By 
choosing an inadmissible fiducial vector, a set of ``coherent states'' spans 
the same space but loses its resolution of unity, and for that reason has
been called a set of weak coherent states. Despite having no resolution of
unity, it is nevertheless shown how the propagator in such a basis may admit
a phase-space path integral representation in essentially the same form as if 
it had a resolution of unity. Our examples are toy models of similar 
situations that arise in current studies of quantum gravity.  

\end{abstract}
\section{Introduction}
The use of coherent states in the construction of coherent state path 
integrals is by now a well-known story \cite{kla1,kla2}.
Central to that construction is the {\it resolution of unity} afforded by a
set of coherent states. By ``weak coherent states'' we mean a set of states 
that are continuously labelled and, for convenience, normalized, but, 
significantly, they
do {\it not} admit a resolution of unity as an appropriately weighted 
integral of one-dimensional projection operators onto the coherent states. 
The example we have in mind is one for which the coherent states are 
defined by a unitary representation of a noncompact group, but the fiducial 
vector is {\it not} chosen from among the dense set of admissible vectors. 
Under circumstances, for example, that the fiducial vector is nevertheless 
a minimum uncertainty state, it becomes possible to define a rigorous 
phase-space path integral representation that is a close analog of 
previously defined cases involving continuous-time regularizations, even for 
these kinds of weak coherent states.

In the present paper, we outline the construction of such a path integral 
for the case of a single affine degree of freedom, i.e., for coherent states 
defined in terms of the {\it affine group} (often called 
the ``$ax+b$ group'') \cite{asl1}. Such coherent states are, up to a 
phase factor \cite{kla3}, equivalent to the more commonly used SU(1,1) 
coherent states. 

It is interesting to note that a field
theory analog of the present phase-space path integral construction, which 
also involves a set of weak
coherent states, arises in a recently developed program aimed at
quantizing the gravitational field \cite{kla4}.
\section{Kinematics and Weak Coherent States}
A suitable basis for the Lie algebra of a one-dimensional affine group 
consists
of the irreducible self-adjoint operators $Q>0$ and $D$ that satisfy the 
affine commutation
relation \cite{asl1} (with $\hbar=1$)
  \bn [Q,D]=iQ\;. \en
The uncertainty product for these operators reads $\Delta Q\Delta 
D\ge\<Q\>/2$, where $\<Q\>\equiv\<\psi|Q|\psi\>, (\Delta Q)^2
\equiv\<(Q-\<Q\>)^2\>$, etc. If $|x\>$,
$0<x<\infty$, denote $\delta$-normalized eigenstates of $Q$, i.e., $Q|x\>=
x|x\>$, then there is a two-parameter family of (real, normalized) minimum
uncertainty states given by
  \bn\eta_{\a,\beta}(x)=N_{\alpha,\beta}\,x^\a\,e^{-\beta x},\hskip1cm
\a>-1/2,\hskip.6cm
\beta>0\;, \en
with $N_{\a,\beta}$ chosen to ensure normalization. For convenience, we 
also set $\<Q\>=1$ which leads to a one-parameter family of minimum 
uncertainty states given by 
  \bn \label{e1}\eta_\beta(x)=N_\beta\,x^{\beta-1/2}\,e^{-\beta x},
\hskip1cm\beta>0\;. \en
In this Schr\"odinger representation, $D=-\half i[x(\d/\d x)+(\d/\d x)x]$.
The affine group of interest is a two-parameter unitary group defined by
  \bn U[p,q]\equiv e^{ipQ}\,e^{-i\ln(q)D},\hskip1cm q>0,\hskip.6cm  p\in 
\mathbb R\;. \en

Next, we choose $|\eta\>$ as a normalized fiducial vector, and define our set
of coherent states as composed of the continuously labelled vectors 
  \bn  |p,q\>\equiv U[p,q]|\eta\>\;,  \en
for all $-\infty<p<\infty$, $0<q<\infty$. The overlap function of two such 
coherent states is given by
  \bn \<p,q|r,s\>=\frac{1}{\sqrt{qs}}\int_0^\infty\eta^*(x/q)\,e^{-ix(p-r)}\,
\eta(x/s)\,dx  \label{e3}\en
for a general fiducial vector. For the specific functional form (\ref{e1}), 
this expression may be evaluated and reads
  \bn\<p,q|r,s\>=\{(qs)^{-1/2}/[\half(q^{-1}+s^{-1})+\half 
i\beta^{-1}(p-r)]\}^{2\beta}\;.  \label{e65}\en
Observe, apart from the factor $(qs)^{-\beta}$, this expression is an
analytic function of $q^{-1}+i\beta^{-1}p$ and $s^{-1}-i\beta^{-1}r$.

Now, we require the usual resolution of unity. Since $dp\,dq$ is the 
(left-) invariant group measure in these coordinates, we anticipate an 
expression of the form
  \bn \label{e66}c\tint\<p,q|r,s\>\<r,s|t,u\>\,dr\,ds=\<p,q|t,u\> \en
for some constant $c$, $0<c<\infty$. Indeed, for the kernels given by 
(\ref{e65}) we find that (\ref{e66}) is valid whenever $\beta>1/2$, in 
which case $c=1/\{2\pi[1-1/(2\beta)]\}$. However, (\ref{e66}) {\it fails if} 
$0<\beta\le1/2$. More generally, an equation such as (\ref{e66}) requires 
the {\it fiducial vector admissibility condition}
  \bn \<Q^{-1}\>=\int_0^\infty\,x^{-1}\,|\eta(x)|^2\,dx<\infty\;.  \en
This condition is evidently false whenever $0<\beta\le1/2$.

The foregoing is a well-known story \cite{asl1}---and, indeed, the same 
issue also enters into
wavelet theory as well \cite{dau}.
\subsection{Positive-definite functions}
The coherent state overlap function (\ref{e3}) clearly fulfills the 
condition that
  \bn \sum_{j,k=1}^N\a^*_j\,\a_k\,\<p_j,q_j|p_k,q_k\>\ge0 \en
holds for all $N<\infty$, and for the special case of (\ref{e65}), this 
important positive-definite-function
condition holds over the whole range of $\beta$, $\beta>0$. We recall that a 
continuous,
positive-definite function may be adopted as a {\it reproducing kernel}, 
which may then be used to define a {\it reproducing kernel Hilbert space} 
\cite{mes}.
In the special examples based on (\ref{e65}), such a space is a functional 
Hilbert space ${\cal C}_\beta$ 
composed of bounded, continuous functions. A dense set of elements in 
${\cal C}_\beta$ is composed of functions of the form
  \bn  \psi(p,q)\equiv\sum_{j=1}^J\,\a_j\,\<p,q|p_j,q_j\>,\hskip1cm 
J<\infty\;. \en
Let another such element be denoted by
  \bn  \phi(p,q)\equiv\sum_{k=1}^K\,\gamma_k\,\<p,q|r_k,s_k\>,\hskip1cm 
K<\infty\;.  \en
In this case, the inner product of these two elements is {\it defined} to be
  \bn  \<\psi|\phi\>\equiv(\psi,\phi)\equiv\sum_{j=1}^J\sum_{k=1}^K\,
\a^*_j\,\gamma_k\,\<p_j,q_j|r_k,s_k\>\;.  \en
The completion of this space obtained by including the limit points of 
all Cauchy sequences in the norm $\|\psi\|\equiv +\sqrt{(\psi,\psi)}$ 
determines the reproducing kernel Hilbert space ${\cal C}_\beta$ in each 
case. It is noteworthy that the
resultant spaces ${\cal C}_\beta$ are mutually disjoint for distinct 
$\beta$ values, $\beta>0$, save for the zero element which is common to 
all ${\cal C}_\beta$.  
  
When $\beta>1/2$ it turns out that there is an {\it alternative} (local 
integral) way to evaluate the inner product of any two elements in 
${\cal C}_\beta$. In that case, the coherent states admit a resolution 
of unity and therefore 
  \bn &&\<\psi|\phi\>=(\psi,\phi)=\tint\psi(p,q)^*\,\phi(p,q)\,d\mu(p,q)
\;,\nonumber\\
 &&\hskip1cm d\mu(p,q)\equiv dp\,dq/2\pi[1-1/(2\beta)],\hskip.6cm\beta>1/2
\;, \en
integrated over the space ${\mathbb R}\times{\mathbb R}^+$.

The resolution of unity can be written in a more abstract form as
  \bn  \tint |p,q\>\<p,q|\,d\mu(p,q)=\one\;,  \en
and this expression is of paramount use in deriving a coherent state path 
integral for the propagator along conventional lines. For comparison 
purposes to what follows,
we recall the most common construction \cite{kla2} given, e.g., by
  \bn &&\<p'',q''|\,e^{-iT\H}\,|p',q'\>=\<p'',q''|\,e^{-i\ep\H}\,\cdots
\,e^{-i\ep\H}\,|p',q'\>\nonumber\\
   &&\hskip1cm=\tint\cdots\tint\Pi_{l=0}^N\<p_{l+1},q_{l+1}|\,e^{-i\ep\H}
\,|p_l,q_l\>\,\Pi_{l=1}^Nd\mu(p_l,q_l)\nonumber\\
&&\hskip1cm=\lim_{\ep\ra0}\tint\cdots\tint\Pi_{l=0}^N\<p_{l+1},q_{l+1}|
(\one-i\ep\H)|p_l,q_l\>\,\Pi_{l=1}^N d\mu(p_l,q_l)\nonumber\\
&&\hskip1cm=\int e^{i\tint[i\<p,q|(d/dt)|p,q\>-\<p,q|\H|p,q\>]\,dt}\,
\D\mu(p,q)\nonumber\\
&&\hskip1cm={\cal M}\int e^{-i\tint[q\,{\dot p}+H(p,q)]\,dt}\,\D p\,
\D q\;, \en
where $\ep\equiv T/(N+1)$ and $-\tint q\,dp$ (and not $\tint p\,dq$) occurs 
because of our phase convention. In this expression, $p_{N+1},q_{N+1}
\equiv p'',q''$; $p_0,q_0\equiv p',q'$, and in the last two lines we have 
(unjustifyably!) interchanged the integrations and the continuum limit and 
written for the integrand the form it would take for continuous and 
differentiable paths. The line before the interchange of limits generally 
offers a well-defined integral representation. Finally, observe that this 
entire construction is premised on the existence of the resolution of 
unity---a relation that fails to hold in the case of weak coherent states.
 
\section{Continuous-Time Regularization}
\subsection{Complex polarization}
For any $\beta>0$, all elements of ${\cal C}_\beta$ satisfy the relation
 \bn B\,\psi(p,q)\equiv[-iq^{-1}\d_p+1+\beta^{-1}q\,\d_q]\,\psi(p,q)=0\;, \en
where $\d_p\equiv \d/\d p$, etc.
In this sense we say that each space ${\cal C}_\beta$ satisfies a {\it 
complex polarization} condition \cite{wood}.
Consequently, the functions in $\C$ also satisfy $A\,\psi(p,q)=0$, where 
$A\equiv \half\beta B^\dagger B$. 

Consider first the case where $\beta>1/2$. In that case, $A$ is a 
nonnegative, self-adjoint operator for which $0$ lies in the discrete 
spectrum and this value is separated from the the rest of the spectrum by a 
nonzero gap \cite{paul}. It follows, for any $T>0$, that
  \bn  \lim_{\nu\ra\infty}\,(e^{-\nu T A})\,\delta(p-p')\,\delta(q-q')=
\Pi(p,q;p',q')\;, \en
where $\Pi(p,q;p',q')$ is the integral kernel of a projection operator onto 
the subspace where $A=0$, that is, onto the space $\C$. But that integral 
kernel is also given by $\{2\pi[1-1/(2\beta)]\}^{-1}\<p,q|p',q'\>$. Finally, 
we observe that $A$ is a second-order differential operator, and by means of 
the Feynman-Kac-Stratonovich formula, we may represent the reproducing 
kernel by a functional integral in the limit that the parameter $\nu\ra
\infty$. Specifically, we obtain the representation given by
 \bn &&\hskip-.5cm\<p'',q''|p',q'\>=\lim_{\nu\ra\infty}{\cal N}_\nu\int 
e^{-i\tint q\,{\dot p}\,dt
  -(1/2\nu)\tint[\beta^{-1}q^2\,{\dot p}^2+\beta q^{-2}\,{\dot q}^2]\,dt}
\,\D p\,\D q\nonumber\\
&&\hskip1.8cm=\lim_{\nu\ra\infty}2\pi[1-1/(2\beta)]\,e^{\nu T/2}\int 
e^{-i\tint q\,dp}\,dW^\nu(p,q)\;,  \en
where the first expression is formal while the the second is well defined 
in terms of a pinned Wiener measure $W^\nu$ [pinned so that $p(T),q(T)=
p'',q''$; $p(0),q(0)=p',q'$] on a two-dimensional space of constant 
negative curvature $R=-2/\beta$, and in which the stochastic integral 
$-\tint q\,dp$ is interpreted in the sense of Stratonovich (midpoint 
rule) \cite{rop}. 

When $0<\beta\le1/2$, the situation changes significantly. In this case, 
$0$ is part of the continuous spectrum of $A$. The operators 
$\exp(-\nu T A)$ still form a semi-group and admit a 
Feynman-Kac-Stratonovich representation. However, as $\nu\ra\infty$, 
we need to isolate the subspace of {\it non}square-integrable functions 
$\C$ that make up the desired reproducing kernel Hilbert space. As may 
be expected, it becomes necessary to {\it rescale} the limiting operation 
to extract the desired set of functions. 

A very simple illustration of the desired procedure may be given with the 
example ${\tilde A}\equiv\half {\tilde B}^2$, where ${\tilde B}=-i\d_x$ is 
defined for all $x$, $-\infty<x<\infty$. Here, too, ${\tilde A}=0$ lies in 
the continuous spectrum of $\tilde A$. The desired space of functions is 
composed of those for which $\psi(x)={\rm const.}$, namely a 
one-dimensional space. We can choose the reproducing kernel for this 
space to be identically one. Initially, we observe that 
  \bn && (e^{-\nu T{\tilde A}})\,\delta(x-x')\bigg|_{x=x''}=
\frac{1}{\sqrt{2\pi\nu T}}\,e^{-\frac{(x''-x')^2}{2\nu T}}\nonumber\\
  &&\hskip4.1cm =\tint\,dw^\nu(x) \;,  \en
where $w^\nu(x)$ denotes a pinned Wiener measure with diffusion constant 
$\nu$ \cite{rop}. Finally, the reproducing kernel of interest is obtained 
by the limit
  \bn && 1=\lim_{\nu\ra\infty}\,e^{-\frac{(x''-x')^2}{2\nu T}}\nonumber\\
&&\hskip.3cm =\lim_{\nu\ra\infty}\,\sqrt{2\pi\nu T}\tint dw^\nu(x)\;,  \en
a procedure which has effectively selected out the space of (nonsquare 
integrable) functions for which $\d\psi(x)/\d x=0$. Observe, in the present 
case, that the proper $\nu$-dependent scaling factor may also be determined 
self-consistently since 
    \bn e^{-\frac{(x''-x')^2}{2\nu T}}=\tint dw^\nu(x)/\tint_{x'=0}^{x''=0}
dw^\nu(x)\;. \en

Returning to the affine group case, we assume that the proper rescaling 
factor may be determined self-consistently as well. (This assumption is a 
generalization of a conjecture of Davies \cite{smi} regarding the long-time 
behavior of heat kernels based on the relevant Laplace-Beltrami operator.) 
With this assumption,
   it only remains to introduce the Feynman-Kac-Stratonovich representation 
which leads \cite{kla4} directly to
 \bn &&\<p'',q''|p',q'\>=\lim_{\nu\ra\infty}K_\nu\,(e^{-\nu T A})\,
\delta(p-p')\,\delta(q-q')\bigg|_{p=p'',\,q=q''}\nonumber\\
&&\hskip2.3cm=\lim_{\nu\ra\infty}{\o{\cal N}}_\nu\int\,e^{-i\tint q\,
{\dot p}\,dt}\,e^{-(1/2\nu)\tint[\beta^{-1}q^2\,{\dot p}^2+\beta q^{-2}\,
{\dot q}^2]\,dt}\,\D p\,\D q\nonumber\\
&&\hskip2.28cm=\lim_{\nu\ra\infty}K_\nu\int\,e^{-i\tint q\,dp}\,dW^\nu(p,q)\;,
\label{e75}\en
where 
  \bn [K_\nu]^{-1}\equiv(e^{-\nu T A})\,\delta(p)\,\delta(q-1)
\bigg|_{p=0,\,q=1}\;. \en
These equations provide the sought-for path integral representation for 
$0<\beta\le1/2$ when the Hamiltonian vanishes.
\subsection{Introduction of dynamics}
When $\beta>1/2$, the procedure to introduce a nonzero Hamiltonian has 
been worked out previously \cite{paul}. In particular, when $\beta>1/2$, 
it follows that
 \bn \label{e20}&&\hskip-.3cm\<p'',q''|e^{-i\H T}|p',q'\>\nonumber\\
&&\hskip.4cm=\lim_{\nu\ra\infty}{\cal N}_\nu\int\,e^{-i\tint[q\,{\dot p}+
h(p,q)]\,dt}
\,e^{-(1/2\nu)\tint[\beta^{-1}q^2\,{\dot p}^2+\beta q^{-2}\,{\dot q}^2]\,dt}
\,\D p\,\D q\nonumber\\
  &&\hskip.4cm=\lim_{\nu\ra\infty}2\pi[1-1/(2\beta)]\,e^{\nu T/2}\int 
e^{-i\tint[q\,dp+h(p,q)\,dt]}\,dW^\nu(p,q)\;,  \en
where $\H$ and $h(p,q)$ are related by
 \bn \label{e21}\H\equiv \tint h(p,q)\,|p,q\>\<p,q|\,d\mu(p,q) \en
or equivalently by
 \bn \label{e22}\<p'',q''|\H|p',q'\>=\tint\<p'',q''|p,q\>\,h(p,q)\,
\<p,q|p',q'\>\,d\mu(p,q)\;. \en
Observe that this phase-space path integral involves $h(p,q)$, which is a 
different symbol associated with the operator $\H$ than the previously 
used symbol $H(p,q)$. These two symbols are connected by
  \bn  H(p',q')=\<p',q'|\H|p',q'\>=\tint|\<p',q'|p,q\>|^2\,h(p,q)\,
d\mu(p,q)\;. \en

Equations (\ref{e20}) and (\ref{e21}) [or (\ref{e22})] provide a 
satisfactory solution for a continuous time regularized, coherent state 
path-integral representation for the propagator when the coherent states 
possess a resolution of unity, i.e., in the present case whenever $\beta>1/2$.

We now turn our attention to the case where $0<\beta\le1/2$ and no 
resolution of unity exists. First, we offer several simple examples to 
show, nevertheless, that a solution to this problem may possibly exist.
Initially, observe that if $h(p,q)=k$, a {\it constant}, then the 
expected path-integral representation [cf.~(\ref{e75}) and (\ref{e20})] 
yields the correct answer even though the standard relations (\ref{e21}) 
or (\ref{e22}) are meaningless!

For the next set of examples, with $R$ and $S$ arbitrary $c$-number 
parameters, we assert that
\bn  \label{e25}&&\<p'',q''|\,e^{-i(RQ+SD)T}\,|p',q'\>=
\<p''e^{ST}+(R/S)(e^{ST}-1),q''e^{-ST}|p',q'\>\nonumber\\
  &&\hskip.5cm=\lim_{\nu\ra\infty}K_\nu
\int_{p',\,q'}^{p''e^{ST}+(R/S)(e^{ST}-1),\,
q''e^{-ST}}\,e^{-i\tint q\,dp}\,dW^\nu(p,q)\nonumber\\
&&\hskip.5cm=\lim_{\nu\ra\infty}K_\nu\int_{p',\,q'}^{p'',\,q''}\,
e^{-i\tint[q\,dp+(rq+spq)\,dt]}\,dW^\nu(p,q)\;.  \en
In this relation, $r(t)$ and $s(t)$, $0\le t\le T$, $T>0$, are any 
smooth functions for which
  \bn {\sf T}\,e^{-i\tint_0^T[r(t)Q+s(t)D]\,dt}\equiv e^{-i(RQ+SD)T}\;, \en
where $\sf T$ denotes the time-ordering operator.
In deriving the last line of (\ref{e25}) we have changed variables in a 
well-defined integral and observed that the associated changes that 
appear in the measure will all effectively disappear when $\nu\ra\infty$. 
In particular, the required change of variables is that given by
 \bn&& p(t)\ra p(t)e^{t\,{\o s}(t)}+[{\o r}(t)/{\o s}(t)]
[e^{t\,{\o s}(t)}-1]\;,\nonumber\\
  &&q(t)\ra q(t)e^{-t\,{\o s}(t)}\;, \nonumber\\
  && t\,{\o s}(t)\equiv \tint_0^t \,s(u)\,du,\hskip.5cm 
t\,{\o r}(t)\equiv \tint_0^t \,r(u)\,du\;,\nonumber\\
  && {\o s}(T)\equiv S\;,\hskip1cm {\o r}(T)\equiv R \;.  \label{e34}\en

Equation (\ref{e25}) serves to evaluate the propagator in the case that 
$\H=RQ+SD$---and by linearity and completeness of the basic operators 
(with $T=1$)
$\exp[-i(RQ+SD)]$, Eq.~(\ref{e25}) implicitly evaluates the propagator for 
a much wider class of Hamiltonians.

More generally, we assert in the case $0<\beta\le1/2$ that the following 
path integral representation holds:
 \bn \label{e40}&&\hskip-.4cm\<p'',q''|e^{-i\H T}|p',q'\>\nonumber\\
&&\hskip.4cm=\lim_{\nu\ra\infty}{\o{\cal N}}_\nu\int\,
e^{-i\tint[q\,{\dot p}+h(p,q)]\,dt}\,
e^{-(1/2\nu)\tint[\beta^{-1}q^2\,{\dot p}^2+\beta 
q^{-2}\,{\dot q}^2]\,dt}\,\D p\,\D q\nonumber\\
  &&\hskip.4cm=\lim_{\nu\ra\infty}K_\nu\int 
e^{-i\tint[q\,dp+h(p,q)\,dt]}\,dW^\nu(p,q)\;,  \label{e45}\en
where, in the present case, the connection between $\H$ and $h(p,q)$ 
is implicitly given by
 \bn \<p'',q''|\H|p',q'\>=\lim_{\nu\ra\infty}K_\nu\int\,
e^{-i\tint q\,dp}\,h(p(u),q(u))\,dW^\nu(p,q) \label{e46}\en
for any $u$, $0<u<T$. Indeed---thanks to analyticity---the diagonal 
matrix elements determine the operator uniquely in the present case, 
and we may therefore assert that
 \bn  &&\hskip-.7cmH(p'',q'')\equiv\<p'',q''|\H|p'',q''\>\nonumber\\
  &&\hskip1.07cm=\lim_{\nu\ra\infty}K_\nu\int\,
e^{-i\oint q\,dp}\,h(p(u),q(u))\,dW^\nu(p,q)\;.  \label{e47}\en

Finally, we conjecture that these relations hold, at least, for the 
class of self-adjoint Hamiltonian operators $\H$ composed of 
semi-bounded polynomials of the basic operators $Q$ and $D$.

\section{Conclusion}
In Eqs.~(\ref{e45}) and (\ref{e46}) [or (\ref{e47})] we have arrived at 
our desired goal of representing the propagator in terms of a 
phase-space path integral that determines the propagator as 
coherent-state matrix elements of the evolution operator for a 
special class of {\it weak} coherent states that do not admit a 
conventional coherent state resolution of unity.
It is also noteworthy that Eqs.~(\ref{e45}) and (\ref{e46}) 
[or (\ref{e47})] also hold when $\beta>1/2$, and thus these equations 
characterize the entire range of $\beta$, $\beta>0$.

Finally, we again remark that in recent studies of quantum gravity 
\cite{kla4}, weak coherent states are used in an essential way. And, 
analogous to the elementary examples presented in the present paper, a 
phase-space functional integral representation is constructed and used 
despite the fact that no conventional coherent state resolution of unity 
exists.
\section*{Dedication}
It is a pleasure to dedicate this article to Martin C.~Gutzwiller. 
Martin has been and continues to be an inspiration to us all in his 
dedication to quality research carried out in a careful way. One may even 
suppose that working at a Major Industrial Laboratory for a good part of 
his scientific career has been of particular value in shaping his life's 
work in such a desirable manner!
\section*{Acknowledgements}
Thanks are expressed to Bernhard Bodmann, Achim Kempf, Yehuda Pinchover, 
and Barry Simon for comments and helpful remarks.


\begin{thebibliography}{99}

\bibitem{kla1}J.R. Klauder, ``The Action Option and the Feynman Quantization 
of Spinor Fields in
Terms of Ordinary $c$-Numbers", Annals of Physics {\bf 11}, 123-168 (1960).

\bibitem{kla2}J.R. Klauder and B.-S. Skagerstam, {\it Coherent States: 
Applications in Physics and Mathematical Physics}, (World Scientific, 
Singapore, 1985).

\bibitem{asl1}E.W. Aslaksen and J.R. Klauder, ``Continuous Representation 
Theory Using the Affine Group", J. Math. Phys. {\bf 10}, 2267-2275 (1969).

\bibitem{kla3}J.R. Klauder, ``Quantization {\it Is} Geometry, After All", 
Annals of Physics {\bf 188}, 120-141 (1988). 

\bibitem{kla4}J.R. Klauder, ``Noncanonical Quantization of Gravity. I. 
Foundations of Affine Quantum Gravity'', J. Math. Phys. {\bf 40}, 5860-5882 
(1999).

\bibitem{dau}I. Daubechies, {\it Ten Lectures on Wavelets}, 
(S.I.A.M. Publishers, Philadelphia, 1992).

\bibitem{mes}H. Meschkowski, {\it Hilbertische R\"aume mit Kernfunktion}, 
(Springer Verlag, Berlin, 1962).

\bibitem{wood}J. Sniatycki, {\it Geometric Quantization and 
Quantum Mechanics}, (Springer-Verlag, New York, 1980).

\bibitem{paul}I. Daubechies, J.R. Klauder, and T. Paul, ``Wiener Measures 
for Path Integrals with Affine Kinematic Variables", J. Math. Phys. 
{\bf 28}, 85-102 (1987).

\bibitem{rop}G. Roepstorff, {\it Path Integral Approach to Quantum Physics}, 
(Springer-Verlag, Berlin, 1996).

\bibitem{smi}E.B. Davies, ``Non-Gaussian Aspects of Heat 
Kernel Behaviour'', J. London Math. Soc. {\bf 55}, 105-125 (1997).

\end{thebibliography}
\end{document}